\documentclass[a4paper,notitlepage]{article}
\usepackage{graphicx} 
\usepackage[margin=1in]{geometry}
\usepackage{placeins}
\usepackage{amsmath}
\usepackage{braket}
\usepackage[subtle]{savetrees}
\title{Strain and Interface Effects on Magnetocrystalline Anisotropy of MnN.}
\author{Robert A. Lawrence$^1$, Matt I. J. Probert$^1$}

\date{\today}

\begin{document}
\maketitle
\begin{abstract}
    Thin film effects on the Magnetocrystalline Anisotropy Energy (MAE) of MnN were studied using density functional theory (DFT). Initially, strain effects on bulk MnN were considered as a proxy for lattice-matching induced strain and a linear relationship between the $c/a$ ratio and the MAE was found. A fundamental explanation for this relationship in terms of the underlying point-group symmetry is given, which we show is applicable to all uniaxial magnetic materials. Strain and charge-transfer effects were then considered for an ultra-thin film. It was found that a Ta seed-layer suppresses the net spin moment on the Mn ions, leading to a reduction of the MAE. Charge transfer is shown to be the cause of this, and hence similar effects may be expected at any magnetic heterostructure interface. 
\end{abstract}

\section{Introduction}

At the heart of many spintronic devices is an interface between a magnetically anisotropic layer and another magnetic system \cite{Flebus2024}. One major field of investigation is van der Waals heterostructures -- where magnetic layers are bound together by the weak van der Waals interaction to form devices. A greater technical challenge to study are the covalently-bonded heterostructures, where the interactions between the layers are significantly stronger, and individual layers cannot be simply ``swapped out'' for an equivalent. Optimising non-van der Waals heterostructure devices, including through material discovery, is currently a significant driver of research activity \cite{VallejoFernandez2021,Miura2022}.

In addition to experimental searches, computational research into magnetic materials provides significant insight into underlying materials properties. As well as providing a financially cheaper route to screen materials, computational techniques also enable counterfactual or experimentally unrealisable configurations to be considered. This in turn enables effects to be separated and the creation of general principles to guide future searches -- both experimental and computational -- for high-performance materials. 

To date, most of the \emph{ab initio} computational effort in magnetocrystalline anisotropy has been concerned with bulk systems of infinite crystals \cite{Szunyogh2009,Staunton2004,Wolloch2021}. While important for finding novel high performance magnetic materials, the results of these simulations tend to agree relatively poorly with experimental equivalents. One potential reason for this is the difference between an infinite bulk crystal and a thin layer of material within a heterostructure. In this paper, we will investigate the effects of common interfacial physics phenomena on the magnetocrystalline anisotropy energy (MAE) of an exemplar material, MnN. MnN was chosen because it has a high MAE for a system that does not contain scarce, expensive elements such as Ir and Pt (even if the MAE is comparatively low to the best alloys containing these elements). It is also an antiferromagnetic material; this makes direct experimental probing of the MAE a major challenge and it is usually inferred by considering its role in exchange bias of ferromagnetic layers with known anisotropy. Accordingly, a deeper understanding of what drives the MAE within MnN-based devices will be particularly valuable for the search for future rare-earth free anisotropic antiferromagnets.

Previous experimental efforts \cite{VallejoFernandez2021,Chang2020,Jensen2023} have identified that Ta is one of the most promising seed layers on which to grow MnN due to the strain-enhanced anisotropy. This pairing is not without its downsides, however, as both Mn and N will readily form a solid solution in Ta \cite{Frost2024} with essentially 0 barrier to N diffusion in particular. This has the result that relatively thick ($\approx 30$ nm) layers \cite{VallejoFernandez2021} of MnN need to be grown for the combination to realise its full potential, limiting its use in nanoscale devices. This is highly suggestive of interfacial effects acting to suppress the magnetocrystalline anisotropy. As commercial use of MnN would require a significantly thinner layer if it is to replace Ir-rich alloys such as IrMn$_x$ \cite{Frost2024}, it is desirable to find alternative seed layers that can offer the same -- or higher -- performance in terms of anisotropy at much reduced layer thicknesses. Understanding how the common interfacial effects in these systems may impact the MAE is therefore a key step to streamlining the search for these new, thinner, high-performance heterostructures.

\subsection{Background Theory}

There are many physical processes that occur when two materials are brought together to form an interface. It will be the aim of this paper to -- as far as is possible -- separate out these variables for independent investigation. This will give a greater clarity into the underlying physical mechanisms that can influence the magnetocrystalline anisotropy, thereby enhancing the transferability of this work so that we can make broad conclusions that will also apply to other materials. For certain variables, this is not achievable, even within a density functional model, due to the intrinsic linking of the underlying physical phenomena. In this section, these well-known effects will briefly be reviewed.

\subsection{Non-van der Waals Heterostructures}

Within the field of magnetism, significant research has been done on the so-called van der Waals heterostructures; that is, weakly interacting monolayers bound together solely through the van der Waals interaction (London forces). Due to the lack of direct chemical bonding in these systems, transfer of electrical current is significantly hindered. Our interest is not in these systems, but rather in chemically (i.e. covalent or metallically bound) bound ones, where the delocalised electronic orbitals shared on either side of the heterojunction open up a range of useful properties. 

\subsection{Disorder}

The simplest models of heterojunctions usually present only atomically sharp interfaces. In practice, growing such interfaces is an enormous technical challenge, with ionic diffusion and various point defects having a significant role \cite{Thrarinsdttir2022}. Disorder caused by mechanisms such as these has many effects on the system, ranging from breaking of local symmetry to charge and spin redistribution, all of which may be expected to negatively impact the effective magnetocrystalline anisotropy \cite{Xu2017}.

Critically, the predictability of the performance of devices based on atomically sharp interfaces has created a significant drive to select systems which tend towards (or even achieve) this limit in experiments \cite{Frost2024}.  Ultimately, the effects present in an atomically sharp interface cannot be ``engineered out'' of a heterostructure device, and must instead be ``engineered with'', requiring a deeper understanding of these last remaining effects. Accordingly, we neglect the effects of disorder in this paper in order to concentrate purely on these final effects without needing to consider their interactions with defect and disorder-related mechanisms.

\subsection{Lattice Matching and Strain}

One of the simplest effects to describe (and easiest to isolate) is the effect of lattice matching. A mismatch between lattice parameters leads to an increase in the energy of the interface. To minimise this surface formation energy, the two materials on either side of the interface deform such that they both tend towards the average lattice parameters of the two materials at the interface. This induces a strain in both materials that is energetically unfavourable compared with their infinite bulk state, but enables more favourable atomic overlap at the boundary. This greater bonding-related stabilisation results in a net reduction of the interfacial energy. An additional factor when growing crystals is that some \emph{relative orientations} of the two crystals have less lattice mismatch than others, leading to different Miller planes being orthogonal to the growth direction, depending on the precise choice of substrate/seed layers in the stack.

In the high-strain limit, it is generally preferable to introduce defects rather than to strain the material too severely -- defect formation has become energetically preferable as the deformation to realise the strain increases. Many defect formation mechanisms are possible and exactly which defects will form is material dependent. Defects also disrupt long-range order and are experimentally known to have a negative effect on magnetocrystalline anisotropy, making them of lesser scientific and commercial interest. Additionally, defect states are significantly more expensive to model, and due to the combination of these  factors we will only consider the small strain limit ($<3\%$).

Within DFT simulations, the strain effects may be added simply by straining the isolated cell in the same way as is found in the real system (i.e. uniaxial vs biaxial vs hydrostatic strain). Many previous studies \cite{Wolloch2021,Zhao2024} have shown strain to be a powerful control for magnetocrystalline anisotropy, with both the magnitude and sign of any uniaxial anisotropy (whether it is easy-axis, hard plane or \emph{vice versa}) being strain dependent. When modelling systems created through epitaxial growth on a seed layer, applied biaxial strain (with the third axis allowed to relax) provides the best match between theory and experimental conditions.

We note that the change of MAE with strain may be qualitatively explained in terms of crystal field theory and symmetry breaking. A classic example of this is bulk Fe \cite{Giannopoulos2015} which, with its body centred cubic (bcc) structure, has octahedral point groups ($O_h$ in Sch\"onflies notation) at each of its atomic centres. This has degenerate $d_{xy}$, $d_{xz}$ and $d_{yz}$ orbitals in a so-called $t_{2g}$ set, and a further $e_g$  set consisting of the $d_{z^2}$ and $d_{x^2-y^2}$ orbitals. Here,only a very small crystal field exists, as the homoatomic basis means that the only deviations from perfect spherical symmetry are weak perturbations due to imperfect screening of the charge of neighbouring ions. This cubic symmetry in turn gives rise to the cubic anisotropy that is well recorded for bulk Fe (the small deviations from uniform charge yields the small MAE -- on the order of $\mu$eV per ion). If Fe is strained either uniaxially or biaxially, then certain symmetry operations are broken, reducing the point group to $D_{4h}$. This has a unique fourfold ($C_4$) rotation axis, which  implies a uniaxial anisotropy. The nature of this axis, magnetically hard or easy, depends on the sign of the strain and is not determined by the point group. In this sense, the qualitative changes to anisotropy induced by strain may be understood through a crystal field theory-like approach, even if the quantitative changes still require a more sophisticated electronic theory technique to predict. 

\begin{figure}
    \centering
    \includegraphics[width=0.5\linewidth]{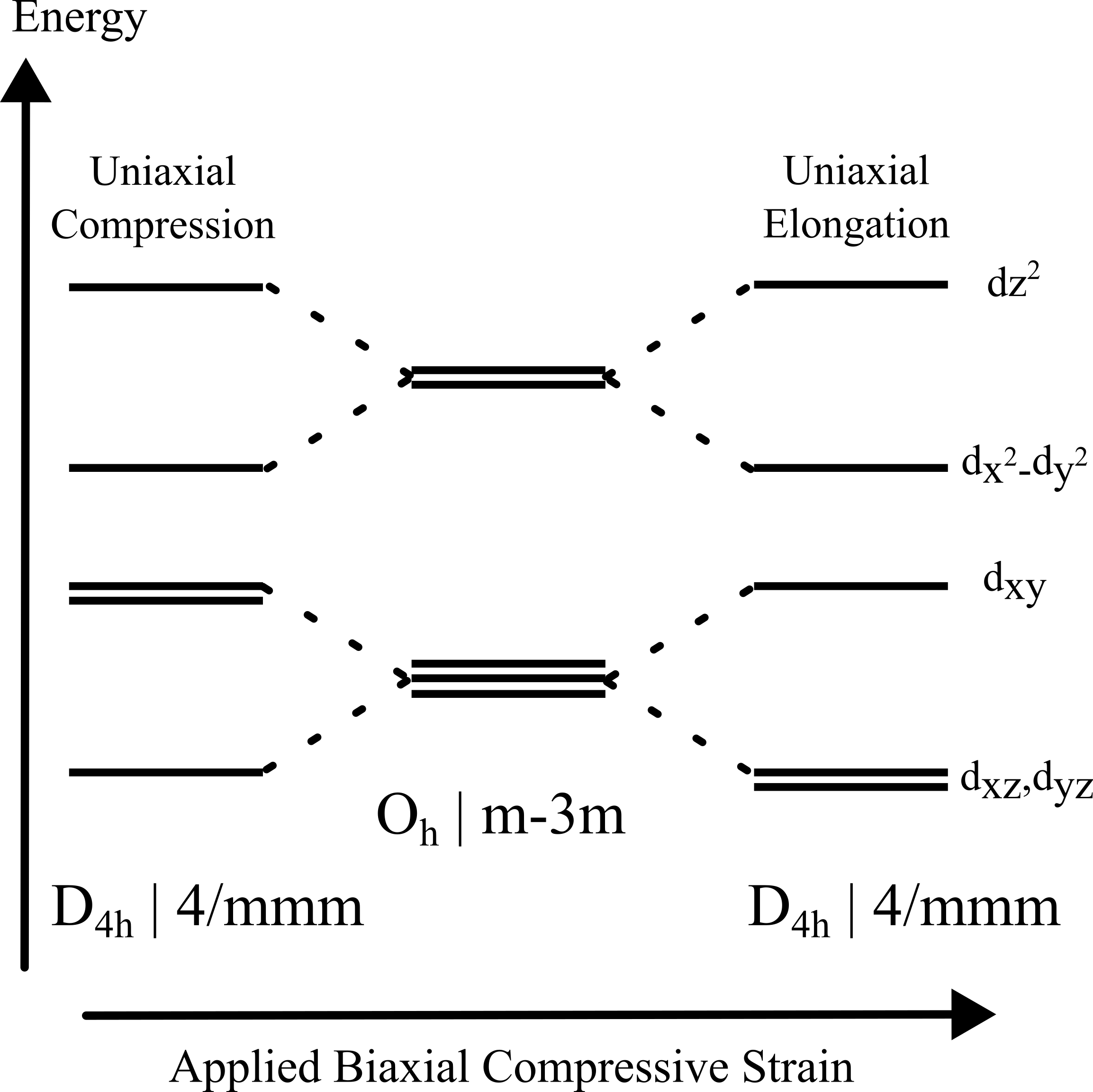}
    \caption{Schematic of changing crystal field splitting against relative uniaxial elongation / compression of an octahedral point group. The magnitude of the splitting depends on the magnitude of the charge asphericity that cause the splitting.}
    \label{fig:D4h-Oh-D4h}
\end{figure}

\subsection{Fermi Level Equilibration and Charge Transfer} 

An additional effect of relevance to interfaces is the redistribution of electrons from one side of the interface to another. This process is driven by Fermi level equilibration \cite{Peljo2016}, which can lead to a net accumulation of charge on one side of the interface, and a depletion of charge on the other. Along with the change in the local charge, one would also expect a change to the local \emph{spin}. This is somewhat harder to predict without \emph{a priori} knowledge, however, as adding an electron to a high-spin complex with greater than or equal to half occupancy of the d- (or f-) subshell will decrease the net spin moment, whereas adding an electron to a system with less than half occupancy will increase the spin moment. A further complication to this is that we are dealing with a quantum mechanical system, and therefore it is the expectation value of spin that matters, which means that partial occupancies are valid.

A useful proxy value for charge redistribution is the atom's electronegativity, which is defined as the ability of an atom to attract an electron towards itself within a covalent bond. For an interface, one may use this alongside \emph{ceteris paribus} thinking to make a rough conclusion about what will happen to the number of electrons at a site; ignoring itinerant (delocalised) magnetism, this is sufficient to determine what will happen to the local spin moments. Within a given material, the MAE is strongly correlated with net moment, which may be altered in a heterostructure due to charge redistribution, as it takes less energy to rotate a smaller spin moment than a larger one. Within DFT simulations, this effect may be quantified by comparing site projections of the spin and charge density, using schemes such as Mulliken decomposition \cite{Mulliken1955}.

\subsection{Shape Anisotropy}

Finally, when comparing infinite bulk MnN with a bilayer system, one must also consider the effects of shape anisotropy. In this, an otherwise magnetically isotropic crystal can gain an anisotropy due to the extent and shape of the physical system. This effect is not present within an infinite bulk crystal \cite{Bruckner2021}, and it is not clear whether a periodically arranged surface would also have this cancellation (as there are an infinite array of surfaces with no ``final surface'' present) or not. To ensure the shape anisotropy is discounted between the different cases in the thin film limit, MAE calculations will be performed involving the same bilayer structure, which should remove changes in shape anisotropy between different systems by controlling the shape anisotropy to always be the same. Thus the difference in anisotropy presented in this work will be magnetocrystalline rather than shape-based. For a real system, shape anisotropy should be considered when making comparisons between the bulk and thin film limits.

\section{Computational Methods}

Density functional theory (DFT) simulations were performed using CASTEP \cite{Clark2005}, a plane-wave pseudopotential based DFT code. A plane-wave cut-off energy of 1800 eV was used in conjunction with a Monkhorst-Pack grid\cite{Monkhorst1976} of 20$\times$20$\times$20 for the bulk MnN and a 20$\times$20$\times$1 grid for the ultra-thin MnN films, with the single k-point representing the aperiodic direction. This ensured an energy convergence of due to the basis of 1$\times 10^{-4}$ eV/atom was met.

For a given self-consistent field (SCF), total energies were converged to 1$\times10^{-8}$ eV/atom. This ensured noise due to the SCF process was orders of magnitude lower than the MAE. Structures were relaxed with a two-point steepest descent algorithm until forces were better than $1\times10^{-2}$ eV/$\text{\AA}$ and the total energy difference between subsequent configurations was better than $2\times10^{-5}$ eV/atom.  

\begin{figure}
    \centering
    \includegraphics[width=\linewidth]{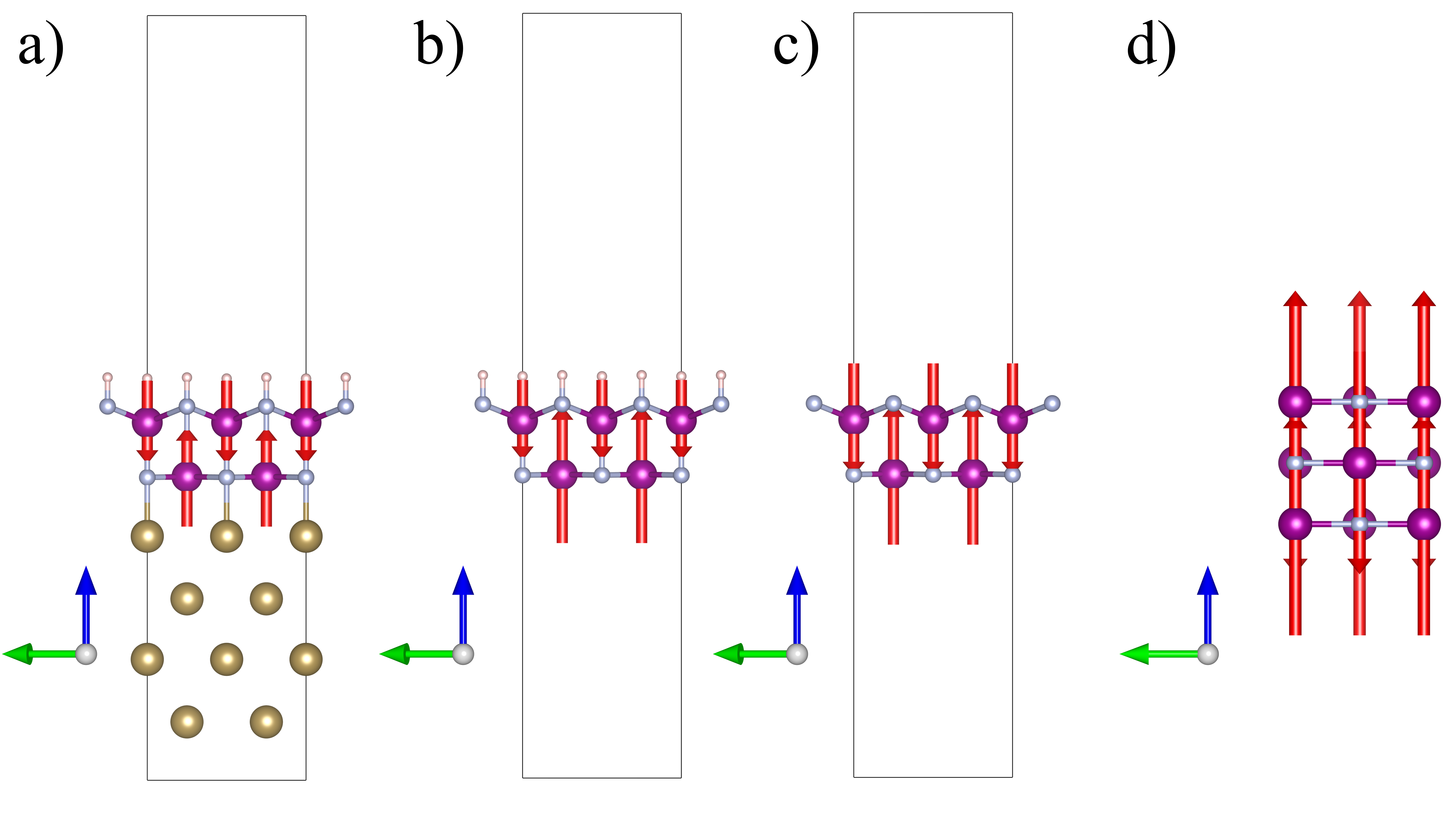}
    \caption{a) H (light pink) -passivated Mn (purple) N (light blue) bilayer on Ta (bronze), b) H-passivated MnN bilayer, c) MnN bilayer, d) Bulk MnN. Spins are represented by red arrows, with the length of the arrow representing the per-site projection of the spin density. Spin magnitudes are reported in table \ref{tab:Spin-length}}
    \label{fig:MnN-Prog}
\end{figure}

To construct unit cells of ultra-thin film MnN, an initial seed layer of [111] Ta was created and relaxed. Then a thin layer of MnN was placed on top under two different translations (with the Mn or N lying directly on top of the Ta) and the whole system was relaxed. In both cases, H-passivation was used to minimise the amount of vacuum gap required to minimise spurious self-interaction of the polar surface with its periodic images, as even the non-polar Ta surface will be affected by the unphysical stray field. The configuration with N on top of Ta (see panel a of figure \ref{fig:MnN-Prog}) was found to be the lowest energy and used thereafter. Finally, to generate extra configurations (and thereby examine the physical mechanisms yielding the thin-film MAE), the Ta and H-passivation layers were removed with no further changes to the unit cell. These are therefore not experimentally realisable systems due to their structurally unsupported nature, but enable direct comparison with the ultra-thin film on Ta (configuration ``a''). 

Requiring the structure not to change also enforced rotating the spin moment without allowing the atoms to relax. For the MAE of the bulk system, the relaxation method outlined in \cite{Magnetochemistry} was used. Relaxation of the ionic degrees of freedom while in the magnetically hard electronic configuration reduced the MAE by 9 $\mu$eV/ion -- indicative of the magnetically excited state corresponding to a fractionally different ground state geometry due to the back-effect of spin-orbit coupling on the lattice. However, this change is essentially negligible, being orders of magnitude smaller than the MAE of bulk MnN. 

The magnetocrystalline anisotropy was evaluated by calculating the difference in total energy calculations for orthogonal configurations, including the experimentally-known easy configuration (along the $\vec{c}$ direction), and all high-symmetry directions within the plane perpendicular to this direction. The difference between the intermediate and easy axes was taken as the MAE; for any kinetic modelling of spin flipping, this represents the peak of the minimum energy pathway and therefore is the experimentally relevant quantity.

\subsection{Surface Reconstruction}

In contrast to the infinite bulk case, in the thin film limit the uppermost layer of N within the MnN bilayer relaxes into the vacuum gap. This is an expected form of surface reconstruction, and is not inhibited by hydrogen passivation. We note that this is an effect caused by our ultra-thin layer of MnN, and would also not be expected for a capped system. Nevertheless, the results we find on the change of MAE between thin-film cases do not depend on the surface reconstruction: the reconstruction is present in every ultra-thin film system as, when conclusions about underlying mechanisms are considered by studying different configurations, the structure including this relaxation is kept constant. 

\section{Results and Discussion}
\FloatBarrier

\subsection{Strain effects on Bulk MnN}\label{ssec:bulk_strain}
One of the most important effects to consider when growing a thin film is lattice matching. To investigate this further, we begin by considering the effects of strain on the MAE of bulk $\theta$-phase MnN, before considering the thin-film limit. 

\begin{figure}
    \centering
    \includegraphics[width=0.7\linewidth]{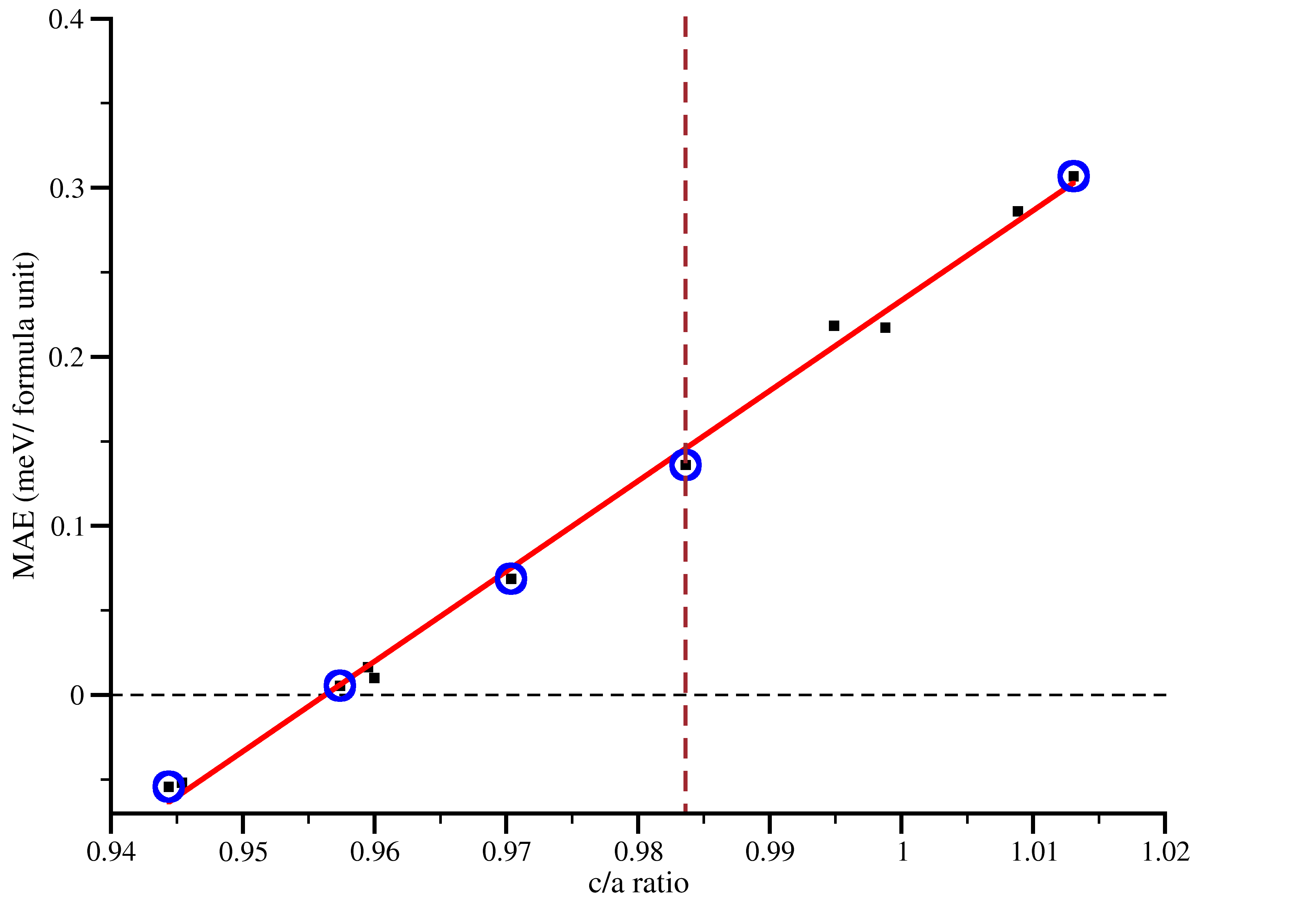}
    \caption{MAE (in meV/ formula unit) against the $c/a$ ratio for bulk phases of MnN. Both the unit cells with relaxed $\vec{c}$ (highlighted by blue circles) and unrelaxed $\vec{c}$ lie on the same line. The vertical dashed line indicate the $c/a$ ratio for bulk MnN, and the dashed horizontal line the cubic anisotropy case.}
    \label{fig:MnN_c2a}
\end{figure}

For each case, a biaxial strain was applied to the $\vec{a},\vec{b}$ lattice vectors, and the MAE was calculated both with and without the relaxation of the $\vec{c}$ lattice vector. For the unrelaxed cases, an arbitrary value of $|\vec{c}|=c \approx 4\, \text{\AA}$ was chosen. The MAE was then plotted against the effective $c/a$ ratio of the system (see figure \ref{fig:MnN_c2a}) and linear regression was performed. The line of best fit for the data is given by
\begin{equation}
    \text{MAE}=  -5.1005+ 5.3338\,\cdot \frac{|\vec{c}|}{|\vec{a}|}\,\,\, \text{meV/formula unit}
\end{equation}
which has a correlation coefficient of 0.998, indicating near-perfect correlation. However, it should be noted that this is only interpolative for small levels of strain ($\pm 3\%)$, and one would expect a deviation from this linearity at higher levels of strain. Nevertheless, this startling linearity regardless of whether the crystal has been geometry-optimised or not suggests a simple underlying physical mechanism.

To explain the linearity, we start by considering the symmetry of the Mn-site. The point group of the ions in MnN is, in Sch\"onflies notation, $D_{4h}$, which is a subgroup of the octahedral ($O_h$) point group, which may be observed in the case of cubic MnN ($c=a$). Both elongation ($c/a>1$) and compression ($c/a<1$) break symmetries (2 of the 4-fold rotation axes) in the $O_h$ point group leading to a reduction in the system's symmetry to $D_{4h}$ regardless of the nature of the strain (i.e. both compressive and tensile strains will lead to the same point group).  We note that other uniaxial $D_{4h}$ systems for which the MAE as a function of strain has been investigated include well-known magnetic materials such as FeCo, FePt and PtMn which also show this same linear trend \cite{Wolloch2021,Magnetochemistry}.

\begin{figure}
    \centering
    \includegraphics[width=0.25\linewidth]{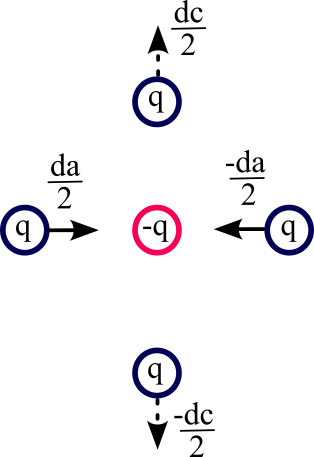}
    \caption{Change in quadrupole moment with application of a biaxial strain. The symmetry of the change in bond lengths with lattice parameters ensures no net dipole moment about the Mn-site, and the change in effective monopole moment with \emph{any} displacement that is not totally symmetric gives rise to a change in the quadrupole moment that is linear with the $c/a$ ratio. Note, in a real system, the effective charge, q, need not be an integer multiple of the electron charge.}
    \label{fig:ca}
\end{figure}

It is instructive to consider a multipole expansion of the so-called crystal field (the electrostatic potential due to the uneven distribution of charge in the system) about the Mn ions. As we change the $c/a$ ratio, we find that we both change the quadrupole moments (which are the only moments that can interact with d-orbitals to lift their mutual degeneracy as they possess the same symmetry) about the Mn atoms and the monopole moment (which provides a rigid shift to the energy that cancels when evaluating MAE due its isotropic nature). For the non-relaxed state, the change in the monopole term of the expansion still leads to a change in the $c$-aligned components of the quadrupole moment. Figure \ref{fig:ca} highlights these motion in a visual schematic.

The effects of these changes on the electronic structure may be seen in figure \ref{fig:D4h-Oh-D4h}, which also notes that for the special case of $c/a=1$ we have the standard octahedral crystal field splitting. For an isolated atom, octahedral splitting would cause cubic anisotropy, such as is observed with bulk Fe, however for binary systems such as MnN, FePt and PtMn, this is evidently not the case. This is due to the fact that in a heteronuclear system the charge imbalances are much higher due to the difference in electronegativities between the species in the system (this even applies in a metallic system -- it controls the degree that the ``sea of electrons'' that comprise metallic bonds \cite{Heine2024} will be localised around an individual nucleus and therefore the local deviations from charge neutrality in the system). For a homonuclear system, there is no difference due to the species, and the remnant charge discrepancy is due to fractionally higher electron density in regions directly between neighbouring atoms (see e.g. \cite{Liu2021}). This deviation from spherical symmetry is very small, and hence the magnitude of the crystal field splitting (and the MAE) is much smaller for BCC Fe than in the example binary systems.

From these simple crystal field arguments, one would anticipate that any system where the atoms have octahedral symmetry would exhibit cubic anisotropy. However, this is not the case (see figure \ref{fig:MnN_c2a}), and the reasoning for this is discussed in further detail in appendix \ref{App:SOC}.

In brief, the spin-orbit coupling is dependent on the Hessian of the scalar potential and therefore it also depends on the electric quadrupole moments, which vary linearly with applied strain. Different spin orientations (along easy or along hard) are affected differently by different quadrupole moments, and therefore while the spin-orbit contribution will vary linearly with strain for each spin configuration, the rate at which it varies is dependent on the spin orientation, leading to a linear variation in MAE. 

In terms of the challenge of device development, it is worth noting that elongating the cell (increasing $c/a$) corresponds to a poorer lattice match -- and accordingly there is a competition between increasing the strain to improve MAE, and generating localised defects which are likely to harm it. Nevertheless, the clear message here is that for MnN a lattice mismatch with compressive in-plane strain will result in much higher performance (in terms of having a high MAE) than a tensile in-plane strain (which could significantly reduce the MAE of the system). We note that simple crystal field arguments indicate that this result will transfer to all $D_{4h}$ systems where the ``non-magnetic'' ion is more electronegative than the ``magnetic'' ion, however this trend will reverse if the relative charges also reverse. 

\FloatBarrier
\subsubsection{Applicability to Other Magnetic Systems}

The linear relationship of strain to MAE has previously been reported for other systems, such as FePt and PtMn \cite{Magnetochemistry}, FeCo \cite{Wolloch2021}, as well as bulk Fe. In common to all of these examples, an applied biaxial strain causes them to undergo a transition from a $D_{4h}^+$ to a $D_{4h}^-$ symmetry, while passing through an $O_h$ configuration (we note that for bulk Fe, the $O_h$ configuration is the ground state BCC structure). This explanation in terms of the underlying point group symmetry for the magnetic ion also explains the underlying trend for the change in anisotropy with respect to strain in these systems. This enables us to state that this will be a common trend in all materials where the point group of the ``magnetic'' ions are all $D_{4h}$ with a common rotation axis. Finally, we note that this also applies to the $O_h$ point group, which is a superset of the $D_{4h}$ point group. The authors also conject that a similar mechanism may also affect the entire $D_{nh}$ class, making this observation of linearity with respect to biaxial strain applicable to most uniaxial magnetic systems. 

\subsection{Strain Effects in the Ultra-thin Film Limit}
\begin{figure}
    \centering
    \includegraphics[width=0.7\linewidth]{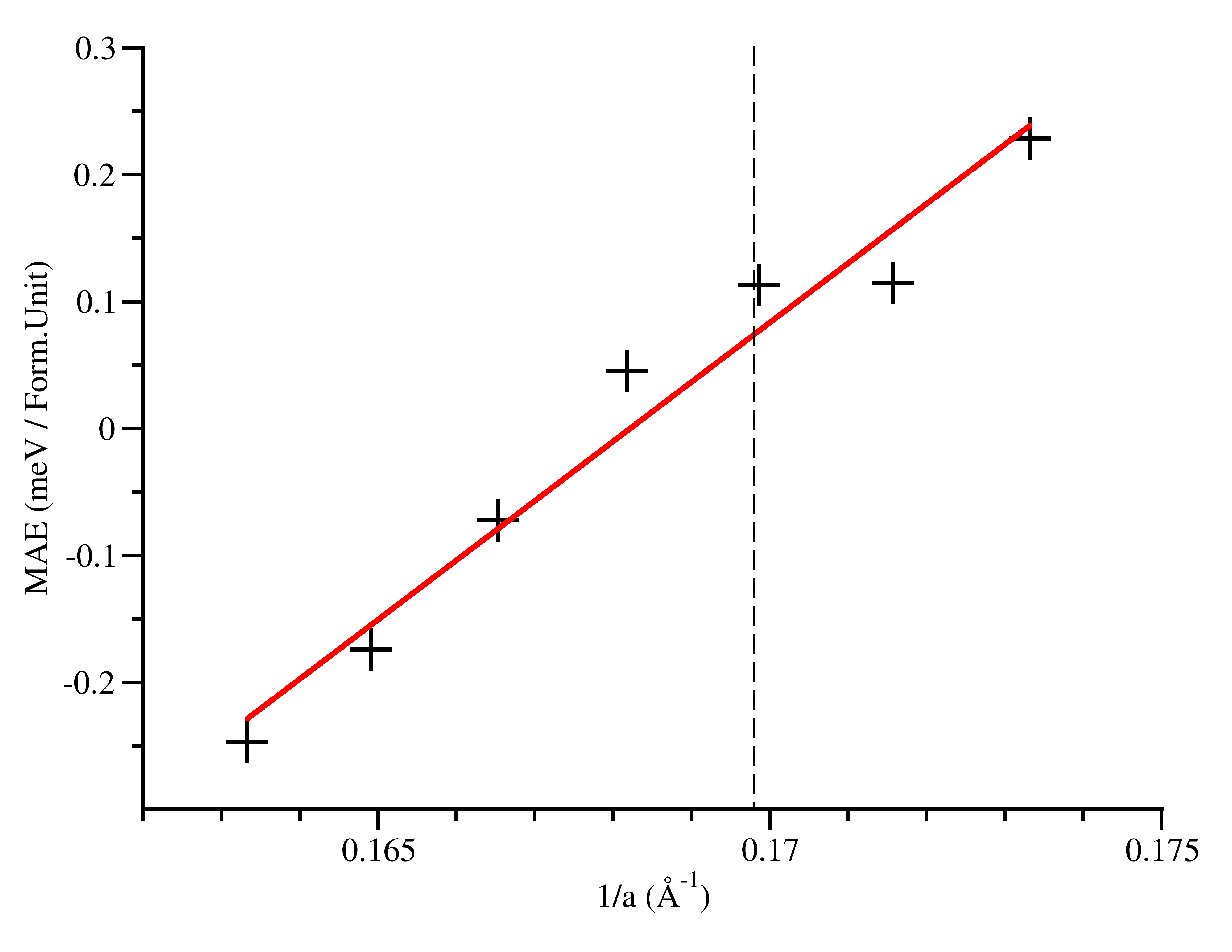}
    \caption{MAE for the ultrathin film case of MnN shown in panel b of figure \ref{fig:MnN-Prog}. The vertical black dotted line indicates the inverse lattice parameter for MnN on a Ta seed layer.}
    \label{fig:MnN_ultrathin}
\end{figure}

\FloatBarrier
Strains were also applied to an ultrathin film of H-passivated MnN (see panel b, figure \ref{fig:MnN-Prog}), and the value of the MAE against the inverse of the $\vec{a}$ lattice constant was recorded. A $c/a$ ratio was not used as the $\vec{c}$ lattice vector was kept fixed throughout these simulations (and is somewhat ill-defined for a thin-film in vacuum), and $\frac{1}{|\vec{a}|}$ is used instead. The results of these are shown in figure \ref{fig:MnN_ultrathin}, with a line of best fit given by   
\begin{equation}
    MAE =\frac{46.81}{|\vec{a}|} -7.874 \text{ meV / formula unit} 
\end{equation}
for MAE in meV/formula unit and $|\vec{a}|=a$ measured in Angstroms. We emphasise that this equation is for the ultra-thin film limit, for a system that is unlikely to be experimentally realisable. Nevertheless, useful information can be drawn about the MAE in such a system that will provide insight into real systems.

We also note that the system underwent a change in-plane, with the elongation of one of the in-plane axes in order to better lattice-match with the Ta. This reduced the symmetry from a $D_{4h}$ to a $D_{2h}$ system, and accordingly an intermediate as well as hard axis was seen in-plane. The intermediate axis (bisecting the bonds) was lower in energy and this is the MAE value reported for these systems, as it is the value relevant to spin flipping events (which are dominated by the lowest-energy saddle point between two potential wells).  This is in agreement with the earlier comments on the crystal field effects applying to all $D_{nh}$ systems, and a linear relationship of MAE is still expected (as seen in figure \ref{fig:MnN_ultrathin}).

\FloatBarrier
\subsection{Non-Strain Effects of a Seed Layer}
\FloatBarrier

However, strain due to lattice matching is not the only effect that occurs when changing from bulk to thin film. Consider the growth of an ultra-thin film of MnN on [111] Ta: although this system may be too thin to represent an experimentally realisable system, it emphasises the subtle effects that may be expected, and makes a qualitative, transferable, analysis of the relevant physical mechanisms simpler.    
To this end, we start by studying an ultra-thin film of MnN grown on a layer of [111] Ta, which is a popular choice for experimental growth of MnN \cite{Dunz2020,VallejoFernandez2021}, with an additional H-passivation layer to suppress the net electric dipole moment and allow for the use of a smaller unit cell (due to the cost of modelling vacuum in a plane-wave basis). The H-passivation is is necessary because the strong difference in electronegativity between N and Mn results in this surface being polar. Initially, two interfacial configurations were chosen -- with either the Mn or N on top of the Ta atoms, and the lower energy configuration following a geometry optimisation (N on top of Ta) was selected for further investigation (see panel a of figure \ref{fig:MnN-Prog}). Then, in order to isolate the effects of the seed layer and H-passivation layers, these layers were removed stepwise with all other parameters kept the same (See panels b and c of figure \ref{fig:MnN-Prog}) and the MAE recalculated. 

As well as inducing an applied strain, the seed layer also chemically interacts with the thin film -- there is a transfer of electrons between the two systems. This effect, which is driven by Fermi level equilibration, leads to a change of the net spin moments in the system and also affects the MAE. This change has multiple origins. Firstly the spin magnitudes may be altered (either quenched or amplified, depending on the precise electronic structure), which affects the MAE. This may readily be seen in equation \ref{eqn:SOC_Ham}, where the MAE is linearly dependent on the spin operator. Additionally, a change of charge will affect the scalar potential, $\Phi$, of the system, thereby changing the effective B-field each electron experiences. Finally, the change in number of electrons on a site will also change which orbitals are occupied -- leading to a change in the number of orbitals $i$ that will contribute to the MAE. 

\begin{table}[]
    \centering
    \begin{tabular}{c|c|c|c}
      Structure   &  S (Mn lower) /$\mu_B$ & S (Mn upper)  /$\mu_B$ & MAE ($\mu$eV/ion)\\
      \hline
       a)  & 1.77 &  -1.48 &  86   \\
       b)  & 2.43 &  -1.42 & 113   \\
       c)  & 2.50 &  -1.99 &  39   \\
       d)  & 3.55 &  -3.55 & 127   \\
    \end{tabular}
    \caption{Simulated spin magnitudes and magnetocrystalline anisotropy energies for the 4 structures presented in figure \ref{fig:MnN-Prog} decomposed by layer. The upper layer has undergone a reconstruction in subfigures (a, b, c) and is additionally H-passivated in subfigures (a and b). Subfigure (a) also features a Ta layer, leading to the suppression of the net spin moment in the lower layer. Negative spin moments indicates spin down, and positive moments indicate spin up.} 
    \label{tab:Spin-length}
\end{table}

Whilst the Heisenberg Hamiltonian for magnetism does not directly include spin magnitude in its description of magnetocrystalline anisotropy 
\begin{equation}
    MAE  = \sum_i K_i\cdot \hat{s_i}^2
\end{equation}
the effect of the changing spin magnitudes are absorbed into the coefficient, $K_i$ through their effect on the energy of the system. Between different systems, however, the spin length is also permitted to change, leading to different anisotropy constants. To demonstrate the effect this might have near an interface, the Mulliken spin projections on the atoms were calculated. These are shown in table \ref{tab:Spin-length}. In this we see that in all cases the magnitude of the spin in the thin-film limit is smaller than that of the bulk case. Given that the exchange bias effect relies most strongly on the interaction of the layers closest to the interfacial boundary, this strongly suggests that the reason for the ``overestimation'' of magnetocrystalline anisotropy energy from density functional theory calculations may be related to the difference between the typically-simulated infinite bulk crystal and the additional effects that are present at heterojunctions. 

The effects of the surface reconstruction may also be seen as the driving mechanism for the asymmetry between the spin magnitudes in the different layers of structure ``c''. However, it is also worth noting that the H-passivation layer quenches the spin it is adjacent to (structure ``b''), and the Ta layer has a still larger quenching effect. It is also worth noting that this quenching is most strongly present in the interfacial layer. This has strong implications for device manufacture, as choosing the capping layer to minimise this quenching will lead to enhanced device performance, as the interfacial atomic layers make the largest contribution to the interaction between adjacent layers in a heterostructure. 

We note that a useful proxy quantity for the degree of charge transfer is electronegativity difference between the ``magnetic'' ion in the active layer and the ions in the seed/capping layers. Maximising performance would require a choice of seed/capping layer that will maximise the individual spin on the magnetic ion, or at least not induce significant quenching at the termination layer. Which direction of electronegativity mismatch will maximise this spin depends on the occupancy of the d (or f) orbitals on the ``magnetic ion'' in the system; where the occupancy is less than half, a less electronegative /higher Fermi level (electron donating) capping layer will be preferable, whereas when the occupancy is more than half a more electronegative / lower Fermi level (electron withdrawing) capping layer will maximise the interaction. 

It is also worth reiterating that the vacuum spacing was not changed when removing the H-passivation layer for configuration ``c''. This potentially leads to the introduction of a dipole field across the system as the screening effects of the H are removed.  This can explain the anomalously low MAE for configuration ``c'' that is not present in either bulk MnN or the H-passivated thin films. In this case, the applied field acts (as in the case of voltage controlled magnetic anisotropy \cite{VCMA}) to modify the anisotropy. Here, we see a reduction in the anisotropy.

From these data, it is possible to conclude that not only lattice matching, but also the degree of charge transfer between layers is important when designing a device where magnetic anisotropy is important. The suppression of spin moments by the seed layer may also explain why high thicknesses of MnN \cite{Dunz2020} are required to make the material perform well; the effects of this suppression will decay to the bulk values as a thicker film is grown. We also note that a ferromagnetic capping layer, commonly used in devices in conjunction with antiferromagnets such as MnN in order to measure its magnetic properties, will have a similar charge-transfer effect. Hence, to maximise device performance, it is necessary to not only select for a desirable lattice match, but also for Fermi-level mismatch to optimise the charge-transfer effect on the MAE. 

\FloatBarrier
\section{Conclusions}
Density Functional Theory was used to study the effects of forming a simple heterostructure on magnetocrystalline anisotropy energy (MAE) for MnN on Ta. The bulk value of MAE for MnN was found to be 127 $\mu$eV/formula unit, reducing to 86 $\mu$eV/formula unit for ultrathin-film MnN on Ta. Strain effects were investigated for the bulk and ultra-thin film limits, and a linear relationship between the $c/a$ ratio and MAE was found with both relaxed and unrelaxed simulation cells obeying the same linear relationship. A simple explanation for this linear relationship in terms of the spin-orbit coupling Hamiltonian was given, and the applicability of this linear trend to all uniaxial systems was posited. 
Within the ultra-thin film limit, the MAE was also found to vary depending on the presence of nearby layers, which was significantly driven by charge transfer induced by Fermi level equilibration leading to spin quenching, especially in the layer immediately forming the interface. These insights may be useful when considering the interactions of neighbouring layers within a magnetic heterostructure for the purpose of improving device performance.

\appendix
\section{Spin Orbit Coupling and Strain Dependence}\label{App:SOC}

In this appendix, we provide a brief derivation of the Spin Orbit Coupling  for an arbitrary field, before considering how the strain induced by lattice matching will affect the MAE in a system such as MnN.

Starting with Maxwell's Equation
\begin{equation}\label{Eqn:MW4}
    \vec{\nabla}\times\vec{B} = \mu_0\left(\vec{J}+\epsilon_0 \frac{d\vec{E}}{dt}\right),
\end{equation}
we may then state that for a crystal in equilibrium with no externally applied current that $\vec{J}=0$. This statement is justified by considering that we are explicitly dealing with systems in equilibrium; there is no time evolution and accordingly we may invoke the principle of detailed balance to show that the current density must be both globally and locally conserved. 

Furthermore, we can consider that for a charged particle passing through a static but spatially varying E-field
\begin{equation}
    \frac{d\vec{E}}{dt} = \frac{d\vec{E}}{d\vec{r}}\cdot\frac{d\vec{r}}{dt} = \nabla{E}\cdot\vec{v}
\end{equation}

We can apply this to an electron of mass $m$ in orbital $\psi(\vec{r})$, using the semi-classical approximation if the spatial variation of $\vec{E}$ is small over the length-scale of $\psi(\vec{r})$, whereupon

\begin{equation}\label{eqn:mybit}
  \frac{d\vec{E}(\vec{r})}{dt} =    \nabla\nabla\Phi(\vec{r})\cdot\psi^*(\vec{r})\frac{i\hbar}{m}\vec{\nabla}\psi(\vec{r}) 
\end{equation}
where $\vec{E}=-\vec{\nabla}\Phi$ and $\nabla\nabla$ represents the Hessian operator. 

We also note that via the Helmholtz theorem, the vector potential $\vec{A}$ may be given by

\begin{equation}\label{eq:A}
    \vec{A}(\vec{r}) = \frac{1}{4\pi}\int \frac{\vec{\nabla}'\times\vec{B}\left(\vec{r}\;\,'\right)}{\vec{r}-\vec{r}\;\,'}d\vec{r}\;\,'
\end{equation}

Finally, returning to Maxwell, we have $\vec{B}=\vec{\nabla}\times\vec{A}$, and using equation \ref{eq:A} with equation \ref{Eqn:MW4} and substituting equation \ref{eqn:mybit}, we can now write an expression for our B-field

\begin{equation}
   \vec{B}(\vec{r})= \vec{\nabla}\times\frac{\mu_0\epsilon_0 i\hbar}{4\pi m}\int\frac{\nabla\nabla\Phi(\vec{r}\;\,')\cdot \psi^*(\vec{r}\;\,')\vec{\nabla}\psi(\vec{r}\;\,')}{\vec{r}-\vec{r}\;\,'}d\vec{r}\;\,'
\end{equation}

We may now write down an expression for $\vec{B}\cdot\vec{S}$ for the magnetic energy of a set of $j$ electrons moving in a static but spatially varying electric field by summing over all single-particle orbitals $\psi_j\left(\vec{r}\right)$,

\begin{equation}\label{eqn:SOC_Ham}
    E_\text{SOC} =  \frac{i\hbar}{4\pi m c^2} \sum_j\int{\psi_j^*\left(\vec{r}\right)}\bigg(\vec{\nabla}\times\int\frac{\nabla\nabla\Phi\left(\vec{r}\;\,'\right)\cdot\psi_j^*(\vec{r}\;\,')\vec{\nabla}\psi_j(\vec{r}\;\,')}{\vec{r}-\vec{r}\;\,'}d\vec{r}\;\,'\bigg) \cdot\vec{S}\psi_j\left(\vec{r}\right) d\vec{r}
\end{equation}
where $\vec{S}$ is the spin operator. We note that this does not reduce to the standard $\vec{L}\cdot\vec{S}$ expression as that is explicitly derived for the case of a spherical potential without any anisotropy; which therefore cannot provide insight into magnetocrystalline anisotropy.

Now, we need to consider how the system will vary with applied strain. In the previous simulations, the applied biaxial strain was \emph{symmetry preserving}. This means that the eigenstates of the initial SOC-free Hamiltonian will remain eigenstates of the final SOC-free Hamiltonian, although their relative energies may change (c.f. figure \ref{fig:D4h-Oh-D4h}).

For sufficiently small strains, the crystal-field perturbation is much smaller than the spin-pairing energy (the energy penalty associated with two electrons occupying the same spatial but opposite spin orbitals). Accordingly, for half-occupied sets of d-orbitals, such as for Mn-atoms in systems where the local moment model applies, we do not expect the occupancy of the d-orbitals to change provided we are in this small-strain limit. For systems that are not half-filled, a sufficiently large change in crystal field splitting will affect these occupancies, however in the zero-temperature limit this requires an actual change of ordering and therefore relatively large changes to the energies of the eigenstates. In the high temperature limit where the Fermi-Dirac function is smoother, the change in orbital occupancy should be present under smaller perturbations. 


We can evaluate the spin-orbit energy (equation \ref{eqn:SOC_Ham}) in terms of effective occupancies of atomic orbitals by projecting our single-particle orbitals onto atomic orbitals $\varphi =\sum_j f_j \psi_j$ (the LCAO approximation), whereupon
\begin{equation}\label{eqn:SOC_braket}
    E_\text{SOC} = \frac{i\hbar}{4\pi mc^2} \sum_j \left|f_j\right|^2\bra{\varphi_j}\vec{\nabla}\times\bra{\varphi\,'_j}\frac{\nabla\nabla\Phi\cdot\vec{\nabla}}{\vec{r}-\vec{r}\;\,'}\ket{\varphi\,'_j}\cdot \vec{S}\ket{\varphi_j}
\end{equation}

Returning to our previous argument, we determined that for a sufficiently small symmetry-preserving strain, the change in orbital energies will be negligible compared with the spin-pairing energy. Hence, the projected occupancy of these d-orbitals, $f_j$, is approximately constant with respect to this strain. The atomic orbitals, being basis functions, are also invariant with respect to strain, as is the gradient of these basis functions and the spin operator applied to them. In this case, the \emph{only} term in equation \ref{eqn:SOC_braket} that will vary with strain is the Hessian of the scalar potential.

We now perform a multipole expansion of our scalar potential about the atomic centre.

\begin{equation}
    \Phi(\vec{r}) = C + {\bf d_i}\cdot \vec{r_i} + {\bf Q_{ij}}\cdot\vec{r}_i\vec{r}_j+{\bf O_{ijk}}\cdot\vec{r}_i\vec{r}_j\vec{r}_k+{\bf H_{ijkl}}\cdot\vec{r}_i\vec{r}_j\vec{r}_k\vec{r}_l+ \dots
\end{equation}
where $C$ is an arbitrary constant that may be set to 0, and the remaining terms are tensors representing the dipole, quadrupole, octupole and hexadecapole terms respectively. We may now consider how the Hessian of $\Phi$ evolves with strain.

The quadrupole moment for a set of discrete charges may be expressed as 
\begin{equation}
    {\bf Q_{ij}} = \sum_l q_l\;\vec{r}_i\vec{r}_j
\end{equation}
 for a set of $l$ discrete charges, $q_l$, and so we can determine how the quadrupole moment varies with applied strain

 \begin{equation}
     \frac{dQ}{d\vec{r}_id\vec{r}_j}=\sum_l \frac{dq_l}{d\vec{r}_id\vec{r}_j}{\vec{r}_i\vec{r}_j}+\frac{dq_l}{d\vec{r}_i}\vec{r}_i+\frac{dq_l}{d\vec{r}_j}\vec{r}_j+q_l
 \end{equation}

From this simplified example it may be seen that providing the change of charge on a site with strain is small, then the quadrupole moment (and therefore our Hessian of our scalar potential) is linear in strain, and lower-order moments will vanish entirely and make no contribution. The contribution of higher-order terms such as octupoles may, by inspection, be seen to be approximately linear in $q_l\cdot\vec{r}_k$ and are therefore not linear in strain. However, we note that since the multipole expansion is a convergent sum, these higher order moments are typically much smaller, and will require relatively large strains to provide a divergence from linearity.

Finally, we consider the symmetry constraints on our system due to the point (site) symmetry. For all crystallographic point groups, there is a leading order non-vanishing moment. These are summarised in table \ref{tab:PGLO}. We note that applied biaxial strain, as is typically relevant to growth conditions, is only symmetry conserving for the point groups which have non-vanishing quadrupole and non-vanishing dipole moments. 

\begin{table}[h]
    \centering
    \begin{tabular}{c|c}
    Leading Moment &  Point Groups \\
    \hline
       Dipole  &  $C_n$, $C_{nv}$ \\
       Quadrupole  & $S_n$, $D_n$, $D_{nd}$, $D_{nh}$\\
       Octupole & $T$, $T_d$\\
       Hexadecapole & $O$, $O_h$, $T_h$ \\
    \end{tabular}
    \caption{Leading order non-vanishing terms of crystallographic point groups}
    \label{tab:PGLO}
\end{table}

Consequently, we expect this result to readily generalise to most crystallographic point groups (we exclude $T$, $T_d$, $O$, $O_h$ and $T_h$ only) provided only that the change in charge with symmetry-conserving atomic displacement is small.

For the excluded groups, we note that the strain induced by lattice-matching is typically symmetry-breaking (for example, $O_h\rightarrow D_{4h}$ on the introduction of a biaxial or uniaxial strain). In this case, we may consider that under these strains, the lower-symmetry point groups will provide a better description of the behaviour of the system over all strain magnitudes, and therefore we would still expect linearity of MAE with respect to strain in a lattice-matching type experiment.



\bibliography{main}
\bibliographystyle{unsrt}
\end{document}